\documentclass{amsart}
\usepackage{graphicx}
\theoremstyle{definition}

\theoremstyle{remark}

\numberwithin{equation}{section}

\begin{document}

\title [Superunification] {A Realistic Superunification}%
\author{J. Towe}
\address{Department of Physics, The Antelope Valley College, Lancaster, CA 93536}%
\email{jtowe@avc.edu}%


\begin{abstract}
It is argued that every quark flavor can be described as a lepton
that has absorbed a superunified field, which is defined. In this
context, a minimal irreducible representation of SUSY SU(5) is
constructed exclusively of the chiral modes that correspond to three
generations of leptons (or quarks) and their scaler superpartners.
The proposed model predicts a new quark--a left-handed (non-strange)
version of the strange quark.
\end{abstract}
\maketitle
\section{Quarks as Superunified Excitations of Leptons}\label{S:intro}%
If one assigns isotopic spins
\begin{equation} \label{E:int}
I_{3}(e^{-}_{L})= -1/2 \end{equation}\\
\begin{equation} \label{E:int}
I_{3}(e^{-}_{R})= 0 \end{equation}\\
and \\
\begin{equation} \label{E:int}I_{3}(\nu^{e^{-}_{L}})=+1/2, \end{equation}
to the light generation of fermions, then the hypercharges of the
light quark generation and the strange quark are given by Y=2Q
-$2I_{3}$:
\begin{equation} \label{E:int}
Y(up)= 2[Q(\nu^{e^{-}})+(2/3)]-2I_{3}(\nu^{e^{-}})=
2[0+(2/3)]-2(1/2)= 1/3
\end{equation}
\begin{equation} \label{E:int}
Y(down) =
2[Q(e^{-}_{L})+(2/3)]-2I_{3}(e^{-}_{L})=2(-1+(2/3))-2(-1/2)=1/3\end{equation}
and \begin{equation} \label{E:int} Y(strange)=
2[Q(e^{-}_{R})+(2/3)]-2I_{3}(e^{-}_{R})=2(-1+(2/3))-2(0)=-2/3,\end{equation}
suggesting that a quark can be interpreted as a lepton that has
absorbed an electrical charge of 2/3.
\par
It will now be demonstrated that transition from a single lepton
to a single quark occurs if the lepton absorbs a gluon and a
spin-2 field that carries a charge of 2/3, a color and a null
isotopic spin. It will also be demonstrated that this interaction
is a consequence of the hetorotic superstring, and specifically of
the graviton vertex operator, which emerges from this version of
the superstring. Because the proposed spin-2 field interfaces the
graviton with electrical charge, color and isotopic spin, this
field is defined as a superunified field.
\par
Let us now consider the herorotic superstring; e.g. let us
consider the gravitino state of momentum k that is described by
the vector-spinor $u^{\mu}$. Graviton emission from this gravitino
ground state is produced by interaction of the gravitino with a
bosonic right-moving (Neveu-Schwarz) prescription, which is
tensored with the fermionic left-moving (Ramond) prescription:
\begin {equation} \label{E:int}
\epsilon_{\mu\nu}[\partial_{\tau}X^{\mu}_{R}(0)+\frac{1}{2}\psi^{\mu}_{R}k
\psi_{R}(0)]\psi^{\nu}_{L}(0)e^{-ikX}
\end{equation}
Expression 1.7 describes what is known as the graviton vertex
operator [J. Bailen, 1994]. The supersymmetric vertex that results
from the above-described interaction is depicted by Figure 1, or
alternatively by Figure 2.
\par
The supersymmetry of the Figure 2 vertex is clearly preserved if
the outgoing spin-(3/2) field is replaced by a fermion-boson pair
of like helicity, as depicted by Figure 3.  Moreover, if the
ingoing fermion in Figure 2 is an electron's neutrino and if the
ingoing boson is a photon, then clearly, the outgoing spin-(2) and
spin-(3/2) fields are just the usual graviton and gravitino; i.e.
then the interaction is a supergravitational interaction. However,
if the ingoing boson in Figure 3 is a gluon, and if the ingoing
fermion is any lepton, then the outgoing fermion is
\textit{necessarily} a quark if the outgoing spin-2 field is a
superunified field (as defined above). This is specifically
established by considering the three possibilities, which are
depicted by Figures 4, 5 and 6. The ingoing fermions in Figures 4,
5 and 6 are respectively the electron's neutrino,
$\nu^{e^{-}}_{L}$, the LH electron, $e^{-}_{L}$ and the RH
electron, $e^{-}_{R}$; while the outgoing fermions are
respectively (based upon the emerging quantum numbers) the up,
down and strange quarks. Thus (given the assignments of isotopic
spin that are indicated by expressions 1.1 through 1.3) the
tree-level, superunified interactions depicted by Figures 4, 5 and
6 clearly require the interpretation of quarks that is indicated
by expressions 1.4 through 1.6. Moreover, because fermionic
generations are indistinguishable at ultra high energies (early in
the cosmological process); i.e. because fermions are massless at
ultra high energies, the proposed supersymmetric theory permits
that the same isotopic spins be assigned to $\mu^{-}_{L}$ and
$\tau^{-}_{L}$ as to $e^{-}_{L}$; to $\nu^{\mu^{-}}_{L}$ and
$\nu^{\tau^{-}}_{L}$ as to $\nu^{e^{-}}_{L}$, and to $\mu^{-}_{R}$
and $\tau^{-}_{R}$ as to $e^{-}_{R}$. In this context, one can
assign hypercharges to those quarks which, as fermions assume
massive status, become the top, bottom and charmed quarks:
\begin{equation} \label{E:int} Y(top)=
2[Q(\nu^{\tau^{-}})+(2/3)]-2I_{3}(\nu^{\tau^{-}})=
2[0+(2/3)]-2(1/2)= 1/3
\end{equation}
\begin{equation} \label{E:int}
Y(bottom) =
2[Q(\tau^{-}_{L})+(2/3)]-2I_{3}(\tau^{-}_{L})=2(-1+(2/3))-2(-1/2)=1/3\end{equation}
and \begin{equation} \label{E:int} Y(charmed)=
2[Q(\nu^{\mu^{-}})+(2/3)]-2I_{3}(\nu^{\mu^{-}})=
2[0+(2/3)]-2(1/2)= 1/3
\end{equation}
Although this result departs from traditional quark theory [D.
Nordstrom, 1992], symmetry appears to permit these assignments of
isotopic spin and hypercharge.
\par
In the above context, one is also motivated to consider the
hypercharge that is generated when $\mu^{-}_{L}$ absorbs a
superunified field:
\begin{equation} \label{E:int}
Y(?) =
2[Q(\mu^{-}_{L})+(2/3)]-2I_{3}(\mu^{-}_{L})=2(-1+(2/3))-2(-1/2)=1/3
\end{equation}
Expression 1.11 clearly indicates a quark that is not currently
recognized. The strange quark is usually regarded as the
generational partner of charmed; but the strange quark cannot be
inserted here (the hypercharge of strange is -2/3). Thus,
expression 1.11 predicts a new quark--a quark that is
characterized by the same quantum numbers that associate with the
strange quark, except strangeness, which is zero for the predicted
quark (the new quark, having an isotopic spin of -1/2, is
left-handed). Note that this prediction cures an anomaly that is
characteristic of the traditional theory of quark generations.
Traditionally the up and top quarks are complemented by
left-handed generational partners: the down and bottom quarks;
while the charmed quark is, according to traditional theory,
complemented only by a right-handed generational partner: the
strange quark. The proposed theory cures this anomaly by providing
a left-handed (non-strange) version of the strange quark. In the
proposed theory, the strange quark is an analogue of the
right-handed electron.
\par
A second consequence of the proposed interpretation of quarks as
superunified excitations of leptons is a SUSY SU(5) model that
precisely accommodates three generations of fermions. This model
will now be considered.
\section{A SUSY SU(5)Model of Three Leptonic Generations}\label{S:intro}
In the context proposed by Section 1, a minimal irreducible
representation $5\oplus10$ of SUSY SU(5) can be precisely
constituted by the chiral modes that associate with the three
leptonic generations and their scaler superpartners. The strictly
leptonic realization of the anti-symmetric 10 = [5,2] of SUSY
SU(5) can be given by
\begin{equation} \label{E:int}
   \mathbf{10_{LEP}}=
   \begin{bmatrix}
    0&e^{-}_{L}&\nu^{e^{-}}_{L}&\tau^{-}_{L}&\nu^{\tau^{-}}_{L}\\
    -e^{-}_{L}&0&e^{-}_{R}&\tau^{-}_{R}&S_{e^{-}}\\
    -\nu^{e^{-}}_{L}&-e^{-}_{R}&0&S_{\tau^{-}}&S_{\nu(\tau)}\\
    -\tau^{-}_{L}&-\tau^{-}_{R}&-S_{\tau^{-}}&0&S_{\nu(e)}\\
    -\nu^{\tau^{-}}_{L}&-S_{e^{-}}&-S_{\nu(\tau)}&-S_{\nu(e)}&0
   \end{bmatrix}
   \end{equation}
where $S_{e^{-}}$ and $S_{\tau^{-}}$ represent the scaler
superpartners of $e^{-}$ and $\tau^{-}$ and where $S_{\nu(e)}$ and
$S_{\nu(\tau)}$ represent the superpartners of the neutrinos. The
strictly leptonic realization of the symmetric 5 = [5,1],
complementing $10_{LEP}$, can be given by
\begin {equation} \label{E:int}
   \mathbf{5_{LEP}}=
   \begin{bmatrix}
    \mu^{-}_{L}\\
    \nu^{\mu^{-}}_{L}\\
    \mu^{-}_{R}\\
    S_{\mu^{-}}\\
    S_{\nu(\mu)}
    \end{bmatrix}
    \end{equation}
where $S_{\mu^{-}}$ is the scaler superpartner of $\mu^{-}$ and
$S_{\nu(\mu)}$ is the superpartner of the muon's neutrino. (Note
that $\mu^{-}_{L}$ and $\mu^{-}_{R}$ correspond to the same scaler
superpartner.) The realization of the anti-symmetric 10=[5,2] in
terms of quarks can be given by
\begin{equation} \label{E:int}
   \mathbf{10_{QRK}}=
   \begin{bmatrix}
   0&d&u&b&t\\
   -d&0&s,S_{B}&s,S_{A}&S_{d}\\
   -u&-s,S_{B}&0&S_{b}&S_{t}&\\
   -b&-s,S_{A}&-S_{b}&0&S_{u}\\
   -t&-S_{d}&-S_{t}&-S_{u}&0
   \end{bmatrix}
   \end{equation}
and the realization in terms of quarks of the symmetric 5 of SU(5)
can be given by
\begin{equation} \label{E:int}
   \mathbf{5_{QRK}}=
   \begin{bmatrix}
   ?\\
   c\\
   s\\
   S_{s}\\
   S_{c}
   \end{bmatrix}
   \end{equation}
where the particle designated '?' represents the new, predicted
quark, where $S_{d},$ $S_{t}$, and $S_{b}$ respectively represent
the scaler superpartners of the down(d), top (t) and bottom (b)
quarks, and where components (s,$S_{A}$) and (s,$S_{B}$)
respectively represent the simultaneous production of a strange
quark and boson that has absorbed a Higgs scaler $S_{A}$; and the
simultaneous production of a strange quark and a boson that has
absorbed a Higgs scaler $S_{B}$. (Note that the charmed quark (c)
and the quark designated by '?' correspond to the same scaler
superpartner.)
\par
The scalers $S_{A}$ and $S_{B}$ are postulated to solve a problem
that is intrinsic to the model under consideration. Specifically,
the interaction that is depicted by Figure 6, which involves an RH
electron and an LH gluon produces a strange quark, whether the
right-handed lepton is an $e^{-}_{R}$, an $\mu^{-}_{R}$ or a
$\tau^{-}_{R}$. It may be however, that this problem can be solved
if one considers the chiral degrees of freedom, described above,
that constitute the quark and lepton realizations of the
irreducible representation $5\oplus10$ of the proposed SUSY SU(5).
The lepton realization of $5\oplus10$ precisely accommodates the
chiral modes that are represented by the three generations of left
and right handed electrons, neutrinos and their scaler
superpartners; but the chiral modes that correspond to the quark
realization of this representation do not exhaust the available
degrees of freedom, unless they include the modes that are
distinguished by the introduction of scaler particles $S_{A}$ and
$S_{B}$---Higgs scalers, in terms of which the above-stated
problem may find a solution. Specifically, the interaction that
produces a strange quark from an $e^{-}_{R}$ (Figure 6) is
understandable in terms of a hypothesis that the outgoing RH boson
has absorbed an anti-scaler $\overline{S}_{A}$ of approximate mass
200 GeV/$c^{2}$ (mass($e^{-})\cong$ .51 GeV/$c^{2}$, mass(s)
$\cong$ 200 GeV/$c^{2}$); and the interaction that produces the
strange quark from a $\tau^{-}_{R}$, as depicted by Figure 7, is
understandable in terms of a hypothesis that the outgoing RH boson
of Figure 7 has absorbed a scaler $S_{B}$ of approximate mass
$10^{3}$ GeV/$c^{2}$ (mass($\tau^{-})\cong$ $10^{5}$ GeV/$c^{2}$,
mass(s) $\cong$ 200 GeV/$c^{2}$). Finally, the outgoing RH boson
of the interaction that is depicted by Figure 8 can be interpreted
as a boson that has not absorbed a scaler (the masses of the
strange quark and the $\mu^{-}_{R}$ are of the same order of
magnitude).
\par
The above hypothesis may also provide an explanation of the two
mass scales $M_{X}$ and $M_{Y}$ that characterize SUSY SU(5)
theories generally. Because the difference $[\Delta M]_{X-Y}$
$\equiv$ $M_{X}$-$M_{Y}\cong10^{15}$ GeV/$c^{2}$ between the mass
scales $M_{X}=10^{18}$ GeV/$c^{2}$ and $M_{Y}=10^{3}$ GeV/$c^{2}$,
is so large that the difference $M_{X}$-$M_{Y}$ appears unrelated
to $M_{A}$-$M_{B}$. But $[\Delta M]_{X-Y}$ and $[\Delta M]_{A-B}$
may represent two states of a running hierarchy that depends upon
the level of energy per particle. In this hypothesis, the sum of
$[\Delta M]_{X-Y}$ and the energy level to which $[\Delta
M]_{X-Y}$ corresponds would always equal $10^{18}$ GeV; e.g.
$[\Delta M]_{X-Y}$ at 1 TeV is about $10^{15}$ GeV/$c^{2}$. In
this context, a lower bound on the energy level where fermions can
be massless can evidently be calculated; e.g. Since $[\Delta
M]_{X-Y}$$\cong10^{5}$ GeV/$c^{2}$ at the energy level where the
interactions depicted by Figure 6 and Figure 7 occur, the energies
at which fermions can be massless must be greater than or equal to
$10^{13}$ GeV. Note that each component of $10_{LEP}$$_{ij}$:
i,j=1,2,3,4,5 is transformed into its counterpart
$10_{QRK}$$_{ij}$ by tree level interactions like those depicted
by Figures 4, 5 and 6, and that each component $5_{LEP}$$_{i}$ is
transformed into its counterpart $5_{QRK}$$_{i}$ by the same
interactions; so that the postulated SUSY SU(5) symmetry is
preserved by the proposed superunified interactions.
\section{Conclusion}\label{S:intro} The graviton vertex operator was
derived as usual from a bosonic, right-moving Neveu-Schwarz
prescription that is tensored with a fermionic, left-moving Ramond
recipe. It was observed that the supersymmetry of the graviton
vertex is preserved if an outgoing gravitino is replaced with a
fermion--boson pair of like helicity. Secondly, it was observed
that if the ingoing fermion is a lepton of given $I_{3}$ (the
third component of isospin), and if the ingoing boson is a gluon,
then the outgoing fermion is a quark of the same isotopic spin if
the outgoing spin-2 field is a superunified field, which was
defined as a spin-2 field of color and of charge 2/3, and null
isotopic spin. In this context, every quark was characterized as a
lepton that has absorbed a superunified field (it was noted that
generations were not distinguished during the very early
cosmological processes, and in this context it was argued that
those quarks, which ultimately constituted the heavier generations
of quarks can also be characterized as leptons that have absorbed
superunified fields). A consequence of this result is that the
chiral modes that correspond to the three leptonic generations and
their superpartners precisely constitute a minimal irreducible
representation $5\oplus10$ of SUSY SU(5). It was shown that this
representation can also be constructed of quarks, and that this
construction predicts a new quark--a left-handed (non-strange)
version of the strange quark.
\par
The construction of $5\oplus10$ in terms of quarks requires that
two initially unused degrees of freedom be accounted for in terms
of two additional scalers. These scalers are interpreted as two
Higgs scalers, and are regarded as distinguishing the productions
of a strange quark from the $e^{-}_{R}$, from the $\mu^{-}_{R}$
and from the $\tau^{-}_{R}$ at the energy level where fermions are
massive. These scalers were also identified with the scalers that
distinguish the X and Y particles of SUSY SU(5) theory.
Specifically, it was postulated that the difference $[\Delta
M]_{X-Y}$ is a function of energy level per particle; e.g. while
$[\Delta M]_{X-Y}$ is approximately $10^{15}$ GeV/$c^{2}$ at a 1
TeV energy level, it was argued that the energy difference
$[\Delta M]_{X-Y}$ is much smaller at a high energy level.
Accordingly, the energy difference $[\Delta M]_{A-B}$$\equiv$
$M_{A}$-$M_{B}$ which distinguishes the production
$e^{-}_{R}$$\rightarrow$$s_{R}$ from
$\tau^{-}_{R}$$\rightarrow$$s_{R}$ and from
$\mu^{-}_{R}$$\rightarrow$$s_{R}$ was identified as equivalent to
the difference $[\Delta M]_{X-Y}$ evaluated at the energy level
below which fermions become massive. In the context of this
hypothesis, it was concluded that $[\Delta M]_{X-Y}$
$\equiv$$10^{13}$ GeV/$c^{2}$ represents a lower bound for
massless fermions. According to the proposed hypothesis, the
difference $[\Delta M]_{X-Y}$ approaches zero as the energy level
approaches $10^{18}$ GeV. \section{List of Figures}\label{S:intro}
1. {Graviton Vertex Operator} \par 2. {Alternative Form of Vertex
Operator} \par 3. {Gravitino Replaced by Fermion-Boson Pair of
Like Helicity} \par 4. {Transition from $\nu^{e^{-}}_{L}$ to Up
Quark} \par 5. {Transition from $e^{-}_{L}$ to Down Quark} \par 6.
{Transition from $e^{-}_{R}$ to Strange Quark} \par 7. {Transition
from $\tau^{-}_{R}$ to Strange Quark} \par 8. {Transition from
$\mu^{-}_{R}$ to Strange Quark}




\end{document}